\documentclass[aps,prb,twocolumn,showpacs]{revtex4-1}

\usepackage{graphicx}

\begin{document}

\title{Supersolidity in the second layer of $para$-H$_2$ adsorbed on graphite
}  

\author{M. C. Gordillo}
\affiliation{Departamento de Sistemas F\'{\i}sicos, Qu\'{\i}micos 
y Naturales, Facultad de Ciencias Experimentales, Universidad Pablo de
Olavide, Carretera de Utrera km 1, E-41013 Sevilla, Spain}

\author{J. Boronat}
\affiliation{Departament de F\'{\i}sica, 
Universitat Polit\`ecnica de Catalunya, 
Campus Nord B4-B5, 08034 Barcelona, Spain}

\date{\today}

\begin{abstract}
We calculated the phase diagram of the second layer of
$para$-H$_2$ adsorbed on graphite using quantum Monte Carlo 
methods. The second layer shows an  incommensurate triangular crystal structure.
By using a symmetric wave  
function, that makes possible molecule exchanges, we observed that this 
nearly two-dimensional crystal shows a finite superfluid density 
around a total density of 0.1650 \AA$^{-2}$. 
The superfluid fraction of this supersolid phase was found to be small, 
0.41$\pm$0.05 \%, but still experimentally accessible.  
\end{abstract}
 

\maketitle
\section{Introduction}

The study of the adsorption of quantum gases on 
graphite is a venerable area of study (see for instance Ref. \onlinecite{colebook} and references therein), dating
back to the 80's and 90's of the past century.
A set of calorimetric, torsional oscillator, and neutron diffraction works 
helped to define the  
the phase diagrams both of the first and second helium 
\cite{grey,grey2,chan,cro,cro2}  and,  to a lesser extent,  molecular hydrogen 
layers
\cite{frei1,frei2,vilches1,cui,doble,vilches2,liu,vilches3} on top of that substrate. 
Those studies have recently been complemented by measurements on 
the adsorption of $^4$He on the outer surface of carbon nanotubes \cite{adrian, 
williams}, introducing curvature effects in the layer formation.

In spite of this long period of intense study, some recent experimental work 
shows that the nature of the second $^4$He layer on 
graphite was not perfectly understood.  Refs.  \onlinecite{B12} and 
\onlinecite{B13} hinted at the existence of novel superhexatic and supersolid phases, 
in addition to the well established translationally invariant and triangular 
incommensurate solid.  The supersolidity claim 
was sustained both by an independent set of torsional oscillator measurements \cite{kim} 
and by a theoretical calculation on the same system \cite{prl2020}.  Both works seem to 
point to a registered two-dimensional (2D) crystal phase with a 
superfluid fraction appreciably above zero.  Ref.  \onlinecite{prl2020} proposed that phase 
to be a 7/12 solid, commensurate with the triangular solid of the first layer. 

Surprisingly, this is not the only work in which $^4$He is proposed to form a supersolid phase. 
Ref.  \onlinecite{prb2011a} shows that a registered $\sqrt 3 \times \sqrt 3$ arrangement on the
first layer adsorbed on graphite had a non-negligible but tiny superfluid fraction of 0.67 \%. 
The same calculation repeated for H$_2$ under the same conditions produced a normal solid.  
On the other hand,  a strictly 2D H$_2$ system was found to be supersolid  in the range
between the spinodal and the equilibrium densities \cite{claudio}.  However, the superfluid 
densities were also quite small.   

$para$-H$_2$ has been deeply studied both theoretically and experimentally 
searching for a new superfluid. Its light mass makes it a priori a good 
candidate that could sum up to the paradigmatic case of helium. However, its 
molecule-molecule interaction is much more attractive than helium, hindering the 
formation of a bulk liquid.  To date, there is only evidence of superfluid 
signal in some spectroscopic studies of small doped H$_2$ 
droplets~\cite{Grebenev,Li}, in agreement with several theoretical 
calculations~\cite{Sindzingre,prb1999,Gordillo-h2,Mezzacapo,Sola,Kwon}. It was also 
shown that a metastable H$_2$ glass has a critical temperature $T_c \simeq 1$ 
K, with a superfluid fraction below 1\% \cite{Oleg}.

In this work, we explore the possibility of the second layer of H$_2$ 
adsorbed on graphite
to be a supersolid.  By a supersolid we mean a system that is not
translationally invariant but that has a non-zero superfluid fraction. 
In this way, it would joint the list of known setups with those characteristics that
today includes, besides quasi two-dimensional $^4$He \cite{B13,kim},  some cold gas arrangements \cite{s1,s2,s3}.
To do so, we solve the Schr\"odinger equation that describes the system 
using the first-principles diffusion Monte Carlo (DMC) method. Our study, 
restricted to the $T=0$ limit, shows that the stable phase corresponding to the 
second layer of H$_2$ grown on top of the first solid layer is an 
incommensurate triangular crystal for all the coverages. Our results for the 
superfluid fraction of this solid show a small density island, around the 
lowest density of the second layer, where its value is different from zero. 
Even if this range is small and the superfluid fraction is predicted to be 
tiny, its value is under reach using modern torsional oscillator 
designs~\cite{kim}.     

\section{Method}

The DMC method is  a 
stochastic technique that
allows us to obtain the  ground state of a zero-temperature system of bosons, as the $para$-H$_2$
molecules considered in this work \cite{borobook}.  DMC solves 
exactly the $N$-body Schr\"odinger equation in imaginary time, within some 
statistical errors. The Hamiltonian of the system under study is
\begin{equation} \label{hamiltonian}
  H = \sum_{i=1}^N  \left[ -\frac{\hbar^2}{2m} \nabla_i^2 +
V_{{\rm ext}}(x_i,y_i,z_i) \right] + \sum_{i<j}^N V_{\rm{H_2-H_2}}
(r_{ij}) \ .
\end{equation}
Here, $x_i$, $y_i$, and $z_i$ are the coordinates of each one of the $N$ H$_2$ 
molecules with mass $m$ located both on the first and second layers.  
Since we considered a full corrugated substrate, we have to sum up 
all the  C-H$_2$ interactions, represented by  $V_{{\rm ext}}(x_i,y_i,z_i)$.  
The functional form of the  C-H$_2$  potential \cite{coleh2}  was the same as in 
a previous work for the same system on graphene 
\cite{yo2013}.  The carbon atoms were located in the graphite crystallographic 
positions in the standard way,  positions that were kept fixed.  
The intermolecular potential $V_{\rm{H_2-H_2}}$ is the Silvera and Goldman 
potential  \cite{silvera},
a staple in these kinds of calculations, that depends only 
on the distance between the center of masses of each pair of para-H$_2$ 
molecules, $r_{ij}$, and not on their relative orientations (spherical 
molecules). Its expression is, in atomic units:  
\begin{eqnarray}
V(r_{ij}) =  \exp(1.713 - 1.5671 r_{ij} - 0.00993 r_{ij}^2) \nonumber \\
- \left( 
\frac{12.14}{r_{ij}^6} + \frac{215.2}{r_{ij}^8} +\frac{4813.9}{r_{ij}^{10}}  - \frac{143.1}{r_{ij}^9} \right) f(r_{ij}). 
\end{eqnarray}
where $f(r_{ij}) = \exp(-(1.28 r_m /r_{ij} -1)^2)$ for $r_{ij} < 1.28 r_m$ and 1 otherwise, with
$r_m$ = 3.44 \AA.

In order to reduce the statistical variance of the simulations and to 
fix the system phase,  DMC uses an initial approximation to 
the many-body wave function which acts as guiding drive along the diffusion 
process.  In the present study we used
\begin{eqnarray}
\Phi({\bf r}_1,\ldots,{\bf r}_N) & = & \Phi_J({\bf r}_1,\ldots,{\bf r}_N)
\Phi_1({\bf r}_1,\ldots,{\bf r}_{N_1})  \nonumber \\
& & \times \ \Phi_2({\bf r}_{N_1+1},\ldots,{\bf r}_N) \ ,
\label{phitot}
\end{eqnarray}
with
\begin{equation}
\Phi_J({\bf r}_1,\ldots,{\bf r}_N) = \prod_{i<j}^{N} \exp \left[-\frac{1}{2}
\left(\frac{b}{r_{ij}} \right)^5 \right] \
\label{sverlet}
\end{equation}
a Bijl-Jastrow wave function that depends on $b$,  a variationally optimized parameter whose
value was found to be 3.195 \AA,  in agreement with a previous calculation in a similar system \cite{yo2013}.  
The expression for $\Phi_1$ was
\begin{eqnarray}
\Phi_1({\bf r}_1, {\bf r}_2, \ldots, {\bf r}_{N_1})  =  
\prod_i^{N_1}  \prod_J^{N_C} \exp \left[ -\frac{1}{2} \left( \frac{b_{{\text C}}}{r_{iJ}} \right)^5 \right] \nonumber \\
 \times \prod_i \exp \{ -c_1 [ (x_i-x_{\text{site},i})^2+ (y_i - 
y_{\text{site},i})^2 ]\}  \nonumber \\
 \times \prod_i^{N_1} \exp (-a_1 (z_i-z_1)^2)  \ ,
\label{phicapa}
\end{eqnarray} 
where $N_1$ is the number of  molecules in the first layer, and $r_{iJ}$ represents the
distance between the center or mass of each molecule, $i$, and each of the
$N_C$ carbon atoms, $J$,  in the graphite crystal. The coordinates  
($x_{\text{site}},y_{\text{site}}$) are the crystallographic
positions of the 2D triangular first layer lattice.  
The variational parameters in Eq. (\ref{phicapa}) were
optimized and found to be the same as the ones in Ref. \onlinecite{yo2013}, i.e.,
$a_1 = 3.06$ \AA$^{-2}$ 
$b_C= 2.3$ \AA, $z_1 = 2.9$ \AA. The parameter $c_1$ was obtained from 
linear interpolation between the values
corresponding to densities in the range 0.08 \AA$^{-2}$ ($c_1=  0.61$
\AA$^{-2}$) and 0.10 \AA$^{-2}$ ($c_1 =  1.38$ \AA$^{-2}$) obtained in a 
previous calculation 
including only the first layer \cite{prb2010}.  If in Eq. (\ref{phitot}) 
we assume $\Phi_2=1$, we are dealing only with the first layer.   

On the other hand,  the study of the second layer is carried out by taking 
 $\Phi_2$ as
\begin{eqnarray}
\Phi_2({\bf r}_{N_1+1},\ldots,{\bf r}_{N})  = 
\prod_i^{N_2}  \prod_J^{N_C} \exp \left[ -\frac{1}{2} \left( \frac{b_{{\text
C}}}{r_{iJ}} \right)^5 \right] \nonumber \\
 \times \prod_{i=1}^{N_2} \left[ \sum_{I=N_1+1}^{N} \exp
\{-c_2 [(x_i-x_{\text{site},I})^2 +
(y_i-y_{\text{site},I})^2] \} \ \right] \nonumber \\
\times \prod_i^{N_1} \exp (-a_2 (z_i-z_2)^2) \ \ \ \ \   \ ,
\end{eqnarray}
$N_2$ being the number of molecules in the second layer ($N=N_1+N_2$), $a_2$ = 
1.52 
\AA$^{-2}$ and
$z_2$ = 6 \AA. The parameter $c_2$ was interpolated  as in 
$\Phi_1$.  The points
($x_{\text{site}},y_{\text{site}}$) are again the crystallographic
positions of a 2D lattice, but now for the second layer.  This symmetrized 
Nosanow wave function allows for possible exchanges in the crystal 
\cite{prl2020}, something 
essential to have superfluidity.  When we wanted to have a second-layer liquid, 
we fixed $c_2$ to zero.  The use of an un-symmetrized trial function similar to 
Eq. \ref{phicapa} to describe the second layer produces always higher  energies 
than the ones obtained by using the above equation. 

All the data presented in this work are the result of the average of 10 
independent Monte Carlo histories for each density and for all of the simulation cells used. 
A history is a set of 1.2 10$^5$ Monte Carlo steps, each step involving the change
in the positions, according to the prescription of the diffusion Monte Carlo 
algorithm \cite{borobook},of all the molecules in each of the 300 walkers (configurations)
needed to describe the systems under consideration.  To increase the number of 
Monte Carlo steps, the number of walkers or the number of independent histories do not
vary the results shown. Of those 1.2 10$^5$ 
Monte Carlo steps, the first 2 10$^4$ were 
ignored in the calculation of averages.  Further increases in the length of the stories did not
improve neither the values nor the error bars of the averages obtained. The number of molecules per configuration oscillated
between 216 and 356, depending on the size of the simulation cell.  That meant 144 and
224 molecules in the first layer, respectively.  
In the solid or supersolid arrangements no vacancies were included. 
To avoid mismatch problems with the incommensurate
arrangements in the second and first layers, both were considered to be at the center a nine-cell 
structure created by replication of the original cell by the vectors defined by their 
length and width.  Only the interactions within a given distance of the molecules in the original
simulation cell  were considered
in the Monte Carlo averages.  We checked that the averages for all the magnitudes presented here 
were independent of the size and shape of simulation cell, what implies that this replicating procedure avoids
any problems derived from the incommensurability of the first and second layers.

\begin{figure}[]
\begin{center}
\includegraphics[width=7.5cm]{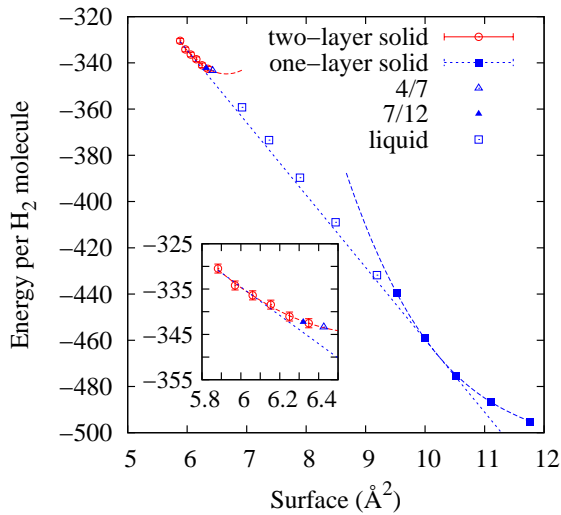}
\caption{Energy per H$_2$ molecule as a function of the inverse of the 
total  density of H$_2$ on graphite.
The dotted line represent the Maxwell construction to determine the transition densities from a one-layer
system to a two-layer arrangement.  The curves are polynomial fits to the data and are intended only as
guides-to-the-eye. The inset is a zoom on the two-layer solid energies.
}
\label{fig1}
\end{center}
\end{figure}

\section{Results}

The first relevant issue is establishing the stability 
limits of the different phases of H$_2$ 
adsorbed on graphite.  This was done to ascertain whether there was
any substantial difference with the phase diagram of the
second layer of H$_2$ on graphene dealt with in Ref. \onlinecite{yo2013}. 
To do so, we calculated the energies per molecule for the different arrangements
we have considered, i.e., a first-layer triangular crystal and a second-layer 
with two possibilities: liquid and a triangular incommensurate 
solid on top of another triangular incommensurate solid in the first layer. 
As indicated above, those results are the average of at least 10 Monte Carlo histories 
for each density. To avoid spurious correlations,  only values located 100 Monte Carlo 
steps away were used to obtain the mean energies.  Our 
DMC  results are shown in Fig.  \ref{fig1}.  
The double-tangent Maxwell construction line indicates that one-layer 
incommensurate solid of 
0.097 $\pm$ 0.002 \AA$^{-2}$ is in equilibrium with two stacked
triangular incommensurate crystals of total density 0.1650 $\pm$ 0.0025 
\AA$^{-2}$, as can be better seen in the inset 
of that figure.  The density of the first layer was 0.100 \AA$^{-2}$,  
density at
which the total energy of the entire arrangement was found to be the lowest 
one. This density translates to  a H$_2$-H$_2$ lattice constant of  
 3.4 \AA. Those results are in good agreement with 
the ones for graphene \cite{yo2013}. Apart from the values 
of the equilibrium densities, the main features of
Fig. \ref{fig1} coincide  with both the graphene calculation and available 
experimental data \cite{doble}, i.e., there are no
stable second-layer liquid  nor 
registered commensurate solids of the second layer with respect the first one 
(either 4/7 or 7/12) since, in both cases, the energies are above the Maxwell 
line.  A picture of those commensurate arrangements can be found in 
Refs. \onlinecite{fukuyama} and \onlinecite{kwon}, respectively.

\begin{figure}[]
\begin{center}
\includegraphics[width=7.5cm]{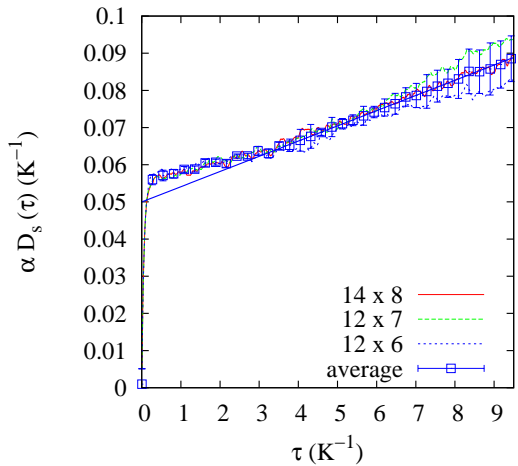}
\caption{Estimator of the superfluid
density for a total density $\rho$ = 0.1650 \AA$^{-2}$ for different sizes of the simulation cell expressed as
multiplets of a single unit cell in the first layer.  Lines, simulation results; open squares, average of all the estimator 
for the three cells displayed.  The straight line represent a linear least-squares fit to the the symbols displayed in the range
(3  $ < \tau <$  8 K$^{-1}$). The slope of that
curve implies a superfluid density of 0.41 $\pm$ 0.05 \%.  
}
\label{fig2}
\end{center}
\end{figure}

The low mass of the H$_2$ molecule makes it a good candidate to exhibit macroscopic quantum behavior. 
To explore this possibility, we studied if we could find a supersolid phase within the stability range of the second-layer triangular solid. 
We chose H$_2$ on top of H$_2$, on top of graphite, because the supersolidity of the first layer had been ruled
out in a previous calculation \cite{prb2011a}.  Following the same procedure as in Ref.  \onlinecite{B13} for
$^4$He adsorbed on graphite, 
we estimated the superfluid fraction in two dimensions $\rho_s/\rho$ of H$_2$ in the second
layer by using the zero-temperature winding number estimator,
\begin{equation} \label{super}
\frac{\rho_s}{\rho}= \lim_{\tau \to \infty} \alpha \left(
\frac{D_s(\tau)}{\tau} \right) \ ,
\end{equation}
with $\tau$ the imaginary time used in the quantum Monte Carlo
simulation. Here, $\alpha = N_2/(4 D_0)$, $D_0 = \hbar^2/(2m)$, and
$D_s(\tau) = \langle [{\bf R}_{CM}(\tau)-{\bf R}_{CM}(0)]^2 \rangle$. ${\bf
R}_{CM}$ is the
position of the center of mass of the $N_2$ H$_2$ molecules located in the second layer.  To perform those calculations, we took into account 
only their $x$ and $y$ coordinates, where periodic boundary conditions apply.  The results obtained for a system with a total density of 0.1650 \AA$^{-2}$ 
are displayed in Fig. \ref{fig2}. There, we show 
$\alpha D_s(\tau)$ vs.  $\tau$,  the superfluid fraction being the slope of the curve for $\tau \to \infty$. As one can see in Fig. \ref{fig2},  $\rho_s/\rho$ is noticeable different 
from zero, i.e., the system is a supersolid at that density.

To be sure that this result did not depend on our chosing of any particular setup,  and to be sure that there were no influence of any possible sluggishness in the superfluid estimator,  we performed DMC simulations using three
different simulation cells.  For instance, a 14 $\times$ 8 simulation cell is made of 14 first-layer unit cells in the $x$ direction, and 
of 8 of those unit cells in the $y$ direction.  Since, as mentioned before, the distance between H$_2$ molecules in that
 first layer  is 3.4 \AA,
this means a  47.6 $\times$ 47.11 \AA$^2$ simulation cell.   Conversely, the dimensions of a 12 $\times$ 6 cell are 
40.8 $\times$ 35.33 \AA$^2$. This implies a surface ratio between both setups of  $\sim$1.55.  The superfluid estimator (\ref{super}) 
was calculated as the average of ten statistically independent simulations and displayed
in Fig. \ref{fig2} as a thin line.  No obvious size trend was found, the values for the three simulation cells being remarkably close to 
each other.  In fact, the open squares in that figure represent the average of the three of them for each $\tau$, the error bars 
corresponding to two standard deviations of that average and including all the values obtained in the simulations. 

To obtain the superfluid fraction,  we performed a least-squares linear fitting procedure to the average of $\alpha D_s(\tau)$ vs. $\tau$ for the three simulation cells considered. 
The $\tau$ values were in the range 
3 $ <\tau < $8 K$^{-1}$.  The result is displayed as the thick continuous line in Fig. \ref{fig2}. The procedure
produces an slope,  corresponding to the superfluid fraction, of 0.41 $\pm$ 0.05 \%. If instead of using the average of $\alpha D_s(\tau)$ as an input in the fitting procedure, we 
consider the values for each simulation cell separately, we get superfluid fractions of 0.36 $\pm$ 0.05 \% for the 12 $\times$ 6 cell, 0.48 $\pm$ 0.05 \% for the 12 $\times$ 7 cell,
and 0.39 $\pm$ 0.05 \% for the 14 $\times$ 7 cell. Larger cells are beyond our calculating capabilities.  
These values are  of the same order as the superfluid 
fraction obtained for the metastable, strictly 2D, systems considered in Ref. \cite{claudio}.  According to the results 
displayed in Fig. \ref{fig1},  the arrangement with a total density of 0.1650 \AA$^{-2}$ is stable,  and thus we can conclude  that this supersolid could be observed. 
We can also see that the least-squares fit also describes well the interval $\tau >$ 8 K$^{-1}$, indicating that no further lengthening of the simulation series is necessary to obtain an accurate value of the superfluid fraction.

\begin{figure}[]
\begin{center}
\includegraphics[width=7.5cm]{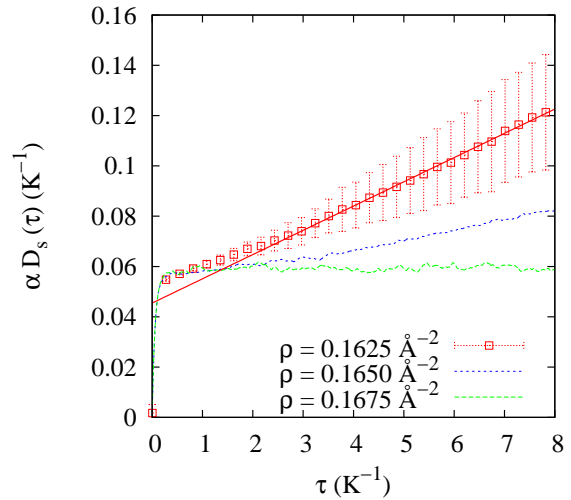}
\caption{Same as in Fig. \ref{fig2}, but for different total 
densities and a fixed 14 $\times$ 8 simulation 
cell. The slope for the density $\rho$ = 0.1625 \AA$^{-2}$ implies a superfluid fraction of 0.96 $\pm$ 0.05 \%. Error bars are only shown for the lower density to make the figure more clear; for the other two densities they are of the same size.}
\label{fig3}
\end{center}
\end{figure}

To check the density range in which that supersolid phase could be stable, we calculated the superfluid fraction (Eq.  \ref{super})
for total densities of 0.1625 \AA$^{-2}$ (metastable) and 0.1675\AA$^{-2}$.  We did not consider the 4/7 and 7/12 unstable phases since their corresponding densities (0.157 and 0.158 \AA$^{-2}$ respectively) are too far away from the stability region.  The results are shown in Fig. \ref{fig3} for a 
14 $\times$ 8 simulation cell.  We observe that, for the lower density, there is an appreciable superfluid fraction of
0.96$\pm$ 0.05 \%.  However, in the other case corresponding to the higher density, the slope of the estimator is zero, implying that the second layer of H$_2$ at that density is a normal solid.

\begin{figure}[]
\begin{center}
\includegraphics[width=7.5cm]{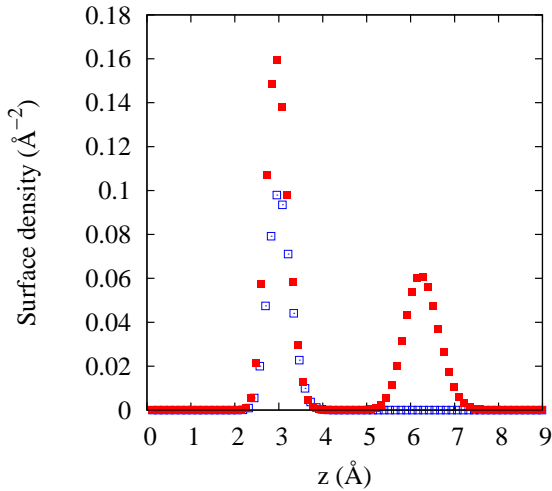}
\caption{Density profile in the direction perpendicular to the 
graphite plane.  Open squares,  density 
corresponding to a first-layer commensurate $\sqrt 3 \times \sqrt 3$ structure; full squares, a two-layer 
system  of total density $\rho$ = 0.1650 \AA$^{-2}$.
}
\label{fig4}
\end{center}
\end{figure}

\section{Conclusions}

Summarizing, our microscopic quantum Monte Carlo approach shows that 
there is a narrow density slice around 0.1650 \AA$^{-2}$ in which 
an incommensurate triangular second-layer H$_2$ supersolid is stable.  We can try to understand the reasons behind this somehow unexpected result.  
The first ingredient is undoubtedly the low density of the crystal considered.  That second-layer density is 0.0650  \AA$^{-2}$.  However, this is larger that the 0.060 \AA$^{-2}$ upper
density limit for which you can see a superfluid fraction in a pure 2D crystal \cite{claudio} and also larger than the 
0.0636\AA$^{-2}$ first layer density for which no superfluid was found in Ref. \onlinecite{prb2011a}. Therefore, an additional factor is needed.  That factor can be understood 
with the help of Fig. \ref{fig4}. There, we show the density profile of
a $\sqrt 3 \times \sqrt 3$  first layer registered phase,  together with the double layer supersolid arrangement in a direction perpendicular ($z$) to the graphite plane.   What we can see is that there is an appreciable difference between the $z$-width of the
first (bare or with H$_2$ on top) and second H$_2$ layers.   In particular, the ratio between their widths, estimated 
considering the first- and second-layer profiles as Gaussians,  is about 1.5.  This implies a considerably larger leeway for the molecules in the second layer to move, and to allow for the exchanges needed to create a superfluid. That is corroborated by the superfluidity fraction of 0.58 $\pm$ 0.05 \% obtained for a second layer of H$_2$ of the same density adsorbed on 
top of a first layer of D$_2$. The second-layer density profile in this case is virtually identical to the presented in Fig. \ref{fig4} and not shown for simplicity. This means that 
the origin of this feature is in the transverse displacement of the molecules in the second layer. 

The only remaining and crucial issue would be about the possibility of detecting such a small superfluid fraction.  In a very recent experimental work \cite{kim}, 
this is shown to be feasible for a double layer of $^4$He on top of graphite, since they were able to detect fractions as small as
0.9\%.  Importantly, the temperatures at which that behavior was seen in $^4$He are well below 0.5 K, what makes us confident that our zero-temperature DMC results for H$_2$ would hold in future experiments to come. Incidentaly, we should say that a very recent calculation \cite{parrinello} points out to the existence of a supersolid in a quite different, but related system, bulk D$_2$ at  
very high pressures.    \\

\begin{acknowledgments} 
We acknowledge financial support from
MCIN/AEI/10.13039/501100011033 (Spain) Grants PID2020-113565GB-C22 and PID2020-113565GB-C21,  from Junta de Andauc\'{\i}a group PAIDI-205,
and the UPO-FEDER grant UPO‐1380159.
J. B.  acknowledges financial support from Secretaria d'Universitats i Recerca del Departament d'Empresa i Coneixement de la Generalitat de Catalunya, co-funded by the European Union Regional Development Fund within the ERDF Operational Program of Catalunya (project QuantumCat, ref. 001-P-001644).
We also acknowledge the use of the C3UPO
computer facilities at the Universidad Pablo de Olavide.
\end{acknowledgments}


\begin{thebibliography}{99}

\bibitem{colebook}  L.W. Bruch, M.W. Cole, and E. Zaremba. Physical Adsorption, Forces and Phenomena
(Dover, New York, 1997).

\bibitem{grey} D. S. Greywall and P. A. Busch, Phys. Rev. Lett. {\bf 67},
3535 (1991).  

\bibitem{grey2} D. S. Greywall, Phys. Rev. B {\bf 47},  309 (1993).

\bibitem{chan} G. Zimmerli, G. Mistura, and M. H. W. Chan, Phys. Rev. Lett.
{\bf 68}, 60 (1992). 

\bibitem{cro} P. A. Crowell and J. D. Reppy, Phys. Rev. Lett. {\bf 70}, 3291 (1993). 

\bibitem{cro2} P. A. Crowell and J. D. Reppy, Phys. Rev. B {\bf 53}, 2701 (1996).

\bibitem{frei1} H. Freimuth and H. Wiechert, Surf. Sci. {\bf 162}, 432 (1985).  

\bibitem{frei2} H. Freimuth and H. Wiechert, Surf. Sci. {\bf 189/190}, 548 (1987).  

\bibitem{vilches1}  J. Ma, D. L. Kingsbury, F. C. Liu, and O. E. Vilches, 
Phys. Rev. Lett. {\bf 61}, 2348 (1988). 
,
\bibitem{cui} J. Cui and S.C. Fain Jr., Phys. Rev. B. {\bf 39} 8628 (1989).

\bibitem{doble} H. Wiechert, in {\em Excitations in Two-Dimensional and Three-Dimensional
Quantum Fluids} edited by A.G.F. Wyatt and H.J. Lauter (Plenum Press, New York, 1991). 
p. 499. 

\bibitem{vilches2} O.E. Vilches, J. Low Temp. Phys. {\bf 89}, 267 (1992). 

\bibitem{liu} W. Liu and S.C. Fain Jr., Phys. Rev. B {\bf 47}, 15965 (1993).  

\bibitem{vilches3} F. C. Liu, Y. M. Liu, and O. E. Vilches, Phys. Rev. B {\bf
51}, 2848 (1995). 

\bibitem{adrian} A. Noury, J. Vergara-Cruz, P. Morfin, B. Placais, M. C. Gordillo, J.
Boronat, S. Balibar, and A. Bachtold, Phys. Rev. Lett \textbf{122}, 165301
(2019).

\bibitem{williams} Emin Menachekanian, Vito Iaia, Mingyu Fan, Jingjing Chen, Chaowei Hu, Ved Mittal, Gengming Liu, Raul Reyes, Fufang Wen, and Gary A. Williams
Phys. Rev. B {\bf 99}, 064503 (2019). 

\bibitem{B12} S. Nakamura, K. Matsui, T. Matsui, and H. Fukuyama,
Phys. Rev. B {\bf 94}, 180501(R) (2016).

\bibitem{B13} J. Nyéki, A. Phillis, A. Ho, D. Lee, P. Coleman, J. Parpia, B.
Cowan, and J. Saunders, Nature Phys. {\bf 13}, 455 (2017).

\bibitem{kim} Jaewon Choi, Alexey  A. Zadorozhko, Jeakyung  Choi, and Eunseong  
Kim,  Phys. Rev. Lett. \textbf{127}, 135301 (2021).

\bibitem{prl2020} M. C. Gordillo and J. Boronat, Phys. Rev. Lett. {\bf 124}, 
205301 (2020).

\bibitem{prb2011a} M. C. Gordillo, C. Cazorla, and J. Boronat, Phys. Rev. B  {\bf 83},
121406(R) (2011).


\bibitem{claudio} C. Cazorla and J. Boronat, Phys. Rev. B {\bf 78}, 134509 (2008). 

\bibitem{Grebenev} S. Grebenev, B. Sartakov, J. P. Toennies, and A. F. Vilesov,
Science {\bf 289}, 1532 (2000). 

\bibitem{Li} H. Li, R. J. Le Roy, P. N. Roy, and A. R. W. McKellar, Phys. Rev. 
Lett. \textbf{105}, 133401 (2010).

\bibitem{Sindzingre} P. Sindzingre, D. M. Ceperley, and M. L. Klein, Phys. Rev. 
Lett. \textbf{67}, 1871 (1991).

\bibitem{prb1999} M.C. Gordillo. Phys. Rev. B {\bf 60} 6790 (1999).

\bibitem{Gordillo-h2} M.C. Gordillo and D.M. Ceperley.  Phys. Rev.  B
{\bf 65}, 174527 (2002).

\bibitem{Mezzacapo} F. Mezzacapo and M. Boninsegni, Phys. Rev. Lett. 
\textbf{100}, 145301  (2008).

\bibitem{Sola} E. Sola and J. Boronat, J. Phys. Chem. A \textbf{115}, 7071 
(2011).

\bibitem{Kwon} Y. Kwon and K.B. Whaley. Phys. Rev. Lett. {\bf 89} 273401 (2002).

\bibitem{Oleg} O. N. Osychenko, R. Rota, and J. Boronat, Phys. Rev. B 
\textbf{85}, 224513 (2012).

\bibitem{s1} J. Léonard, A. Morales, P. Zupancic, T. Esslinger, and T. Donner,
Nature (London) {\bf 543}, 87 (2017),

\bibitem{s2}  J. Li, J. Lee, W. Huang, S. Burchesky, B. Shteynas, F. Top, A.
Jamison, and W. Ketterle,  Nature (London) {\bf 543}, 91
(2017),

\bibitem{s3} L. Tanzi, E. Lucioni, F. Fama, J. Catani, A. Fioretti, C.
Gabbanini, R.N. Bisset, L. Santos, and G. Modugno, Phys. Rev.
Lett. {\bf 122,}  130405 (2019).



\bibitem{borobook} J. Boronat, in \textit{Microscopic Approaches to Quantum
Liquids in Confined Geometries}, Vol. \textbf{4}, ed. E. Krotscheck and J.
Navarro (World Scientific, Singapore,  2002).

\bibitem{coleh2} G. Stan and M. W. Cole, J. Low Temp. Phys. {\bf 110}, 539 (1998).

\bibitem{yo2013} M. C.  Gordillo and J. Boronat. Phys. Rev. B {\bf 87}, 
165403 (2013).


\bibitem{silvera} I. F. Silvera and V. V. Goldman, J. Chem. Phys. {\bf 69}, 4209
(1978).

\bibitem{prb2010} M.C. Gordillo and J. Boronat. Phys. Rev. B
{\bf 81}, 155435 (2010).

\bibitem{fukuyama} H. Fuyuyama.  J. Phys. Soc. Japan {\bf 77} 111013 (2008). 

\bibitem{kwon} Y. Kwon and D.M. Ceperley. Phys. Rev. B {\bf 85} 224501 (2012).

\bibitem{parrinello} C. W. Myung ,B. Hirshberg, and M. Parrinello 
Phys. Rev. Lett. {\bf 128} 045301 (2022). 



\end{thebibliography}
\end{document}